\documentclass[11pt]{article}
\usepackage{epsfig}
\begin{document}
\begin{flushright}SJSU/TP-00-22\\July 2000\end{flushright}
\vspace{1.7in}
\begin{center}\Large{\bf Property Attribution and the Projection
   Postulate \\in Relativistic Quantum Theory}\\
\vspace{1cm}
\normalsize\ J. Finkelstein\footnote[1]{
        Participating Guest, Lawrence Berkeley National Laboratory\\
        \hspace*{\parindent}\hspace*{1em}
        e-mail: JLFINKELSTEIN@lbl.gov}\\
        Department of Physics\\
        San Jos\'{e} State University\\San Jos\'{e}, CA 95192, U.S.A
\end{center}
\begin{abstract}
State-vectors resulting from collapse along the forward light 
cone from a measurement interaction can be used for the attribution
of both local and non-local properties.
\end{abstract}
\newpage
There has recently (see refs.\ \cite{G1} - \cite{Gis1}) been renewed
interest in the question (discussed a while ago in refs.\ \cite{Block} -
\cite{AA3}) of how, in a relativistic theory, a quantum state-vector  
should be assumed to collapse when a measurement is performed.
Of course, even in a non-relativistic theory
the postulate of state-vector collapse upon measurement is, at best,
controversial;
see, for example, ref.\ \cite{Bell}.  In this note, I do not
wish to enter into that controversy;  since I wish to explore the
{\em additional} burden imposed on the  projection postulate by the
lack of a Lorentz-invariant concept of simultaneity, I will in this
note naively accept that postulate, at least in a non-relativistic
context.  Furthermore, I will for the purpose of this note accept
as correct the standard quantum expression for the probabilities of results
of any measurement (which does not require
any assumption of state-vector collapse; see ref.\ \cite{Wig}). 
Thus I will not be considering possible modifications
of that standard expression, as was suggested for example in refs.\
\cite{SS} and \cite{Gis2}.

In this note I will, first, discuss the question of
whether, and how, an assumption of state-vector collapse in a 
relativistic theory can be made which is 
compatible with the standard (``correct'')
expression. I will argue that this can be done in more than one way,
and point out that both  the assumption of collapse along
the backward light cone of the measurement point \cite{HK} and the
assumption of collapse along the forward light cone \cite{me,Mos}
can be considered as special cases of the suggestion \cite{AA3}
that a state-vector can be associated with {\em any} space-like
surface.  Second, I will discuss the suggestion \cite{G1,G3}
that state-vectors associated with particular surfaces  be
used for property attribution. I will present an example in which this
suggestion leads to a rather curious combination of property attributions,
and propose a slight modification of this suggestion for which this
curious feature does not occur.  Finally, I will comment on some 
recently-reported experimental results \cite{Gis1,Gis2} for which the 
preceding considerations are relevant.
 
Hellwig and Kraus (ref.\ \cite{HK}, hereafter called HK) suggested that
a measurement at a space-time point $M$ should cause the state-vector
to collapse along the backward light-cone of point $M$; I will 
refer to this suggestion as ``the HK prescription''. This 
suggestion was criticized by Aharonov and Albert \cite {AA1,AA2}; 
this criticism was elaborated upon by Breuer and
Petruccione \cite {BP} and by Ghirardi \cite {G2}.  It is clear
that the HK prescription does reproduce the standard \cite{Wig}  answer 
for a measurement of any local observable; however, it can be seen that
this prescription would fail if it were possible to measure arbitrary
non-local observables.  One might attempt to defend the HK prescription on
the basis of the result claimed by Landau and Peierls \cite {LP} 
that no non-local observable can be measured.  However, Aharonov and
Albert \cite {AA1} showed that it {\em is} possible to measure
{\em certain} non-local observables (in a sense to be discussed below). 
While this result invalidates the
defense of HK based on the Landau-Peierls claim, it does not yet settle
the question of the viability of the HK prescription, unless it is shown that
this prescription fails for those special non-local observables which in fact 
can be measured.

  Ref.\ \cite {BP} states
``Hellwig and Kraus have shown their prescription \ldots leads to the
correct predictions for all probabilities of {\em local} measurements.
However, this state vector reduction prescription fails for the 
measurement of nonlocal observables''. From this statement, one might
get the impression that the HK prescription can give {\em incorrect}
predictions (or at least fails to give correct predictions) for 
measurements of non-local observables. 
In this note I will not be concerned with any ontological implications
that anyone may or may not wish to draw from the HK prescription,
(or from any other prescription) but I do wish to discuss whether the 
HK prescription does fail to give correct calculated values. 
I begin by summarizing the non-local measurement originally devised
by Aharonov and Albert \cite{AA1} in the streamlined version
recently given  by Ghirardi
\cite{G2}, and will then see whether this measurement can be correctly
described by the HK prescription.  
Two particles, which I will call $A$ and $B$ and which  
each have isotopic spin $I=\frac{1}{2}$, are produced at a common point
and then allowed to separate.  The world-lines of these particles are 
indicated, in some particular reference frame, in fig.\ 1; particle $A$
moves to the left, and particle $B$ moves to the right.  I will let
$|\Psi \rangle _{AB}$ denote the initial state of these two particles.
\begin{figure}
\centering
\epsfig{figure=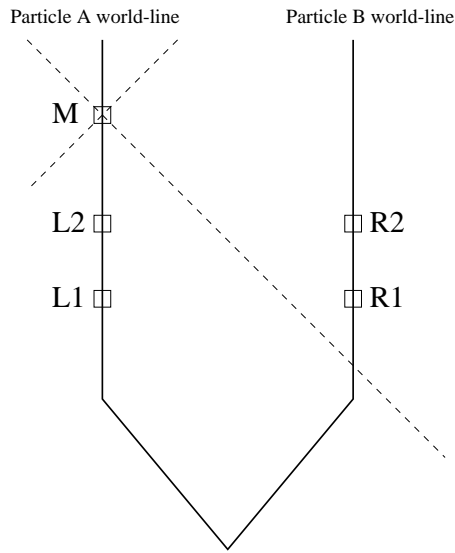}
\caption{``Non-local measurement'' of total isospin at space-time points
    $L1$ and $R1$, and again at $L2$ and $R2$, and local measurement of
    $I_z$ of $A$ at $M$.  Dotted lines represent the light-cone with
    vertex at $M$.}
\end{figure}

It is desired to measure the value of the total isospin of this
two-particle system.  For this purpose, six ``probe particles''
are produced in a particular initial state (given in eq.\ 3.13 of
\cite{G2}) which I will denote by $|\Phi \rangle _1$.  Three of the
probe particles interact with $A$ at the space-time point 
labelled $L1$ in fig.\ 1,
the other three interact with $B$ at the point $R1$. These interactions
are described by a unitary operator I will denote by $U_1$; all of the
details are carefully specified in ref.\ \cite{G2}, but for the present
discussion I need only note a few properties of the specified interaction.
If $ |\Psi \rangle _{AB}$ is the isosinglet state, then $U_1$ acts trivially,
that is
\begin{equation}
U_{1}|I=0,I_{z}=0\rangle _{AB} \otimes |\Phi \rangle _{1} =  
   |I=0,I_{z}=0\rangle _{AB} \otimes |\Phi \rangle _{1}.
\label{Usinglet}\end{equation}
If $ |\Psi \rangle _{AB}$ is any of the isotriplet states, then the
action of $U$ (given explicitly in eqs.\ 3.16--3.18 of \cite{G2})
can be written
\begin{equation}
U|I=1,I_{z}\rangle _{AB}\otimes |\Phi \rangle _{1} = 
    \sum_{I_{z}^{\prime }=-1}^{+1}C_{I_{z},I_{z}^{\prime }}
    |I=1,I_{z}^{\prime }\rangle _{AB}\otimes 
    |\hat{\Phi }_{I_{z},I_{z}^{\prime }}\rangle _{1},
\label{Utriplet}\end{equation}
where the $C_{I_{z},I_{z}}$ are numerical coefficients and where each
of the states $|\hat{\Phi }_{I_{z},I_{z}^{\prime }}\rangle _{1}$
is orthogonal to $|\Phi \rangle _{1}$.

It is important that the state $|\Phi \rangle _{1}$ can be distinguished
from any of the states $|\hat{\Phi }_{I_{z},I_{z}^{\prime }}\rangle _{1}$
by subsequent measurements done on the individual probe particles.
This makes it possible to say, assuming that these subsequent 
measurements are in fact performed, that the state  
$|\Phi \rangle _{1}$ is (or else is not) found. It then follows,
from eqs.\ \ref{Usinglet} and \ref{Utriplet} that
\begin{description}
\item[$\alpha$ ] If the initial state is the isosinglet  
    $ |I=0,I_{z}=0\rangle _{AB}$, then $|\Phi \rangle _{1}$ will surely be
    found.
\item[$\beta$ ] Now assume that a second, similar, measurement is also 
       performed, with six new probe particles in an initial state 
       $|\Phi \rangle _{2}$ (analogous to $|\Phi \rangle _{1}$)
       at the points labelled $L2$ and $R2$ in fig.\ 1.  
       Then, for arbitrary initial
       state of the $AB$ system, if $|\Phi \rangle _{1}$ is in fact
       found, then with probability one $|\Phi \rangle _{2}$ will
       also be found.
\end{description}
These two features, especially feature $\beta $, suggest that we
may regard the observation of $|\Phi \rangle _{1}$ as a preparation
of the $AB$ system in the isosinglet state, and the subsequent
observation of $|\Phi \rangle _{2}$ as a verification of that state.
Although this procedure may {\em result} in a measurement
(at least in the sense of features $\alpha $ and $\beta $) of the
non-local observable $I^2$, it is important to realize that it
{\em consists} of purely-local interactions and measurements.

Now suppose that, at the point labelled $M$ in fig.\ 1, a measurement
of $I_z$ of particle $A$ is performed.  Clearly this does not change 
the preceding
discussion in any way, and in particular features $\alpha $ and $\beta $
are still present. Let $U_2$ denote the unitary operator  for the
interactions at $L2$ and $R2$ (with properties analogous to $U_1$),
and let $P_M$ be the projection operator onto whichever state of
$A$ is in fact found in the measurement at $M$. Then after the 
interactions at $L1$, $L2$, $R1$, and $R2$, and this measurement
at $M$, the joint state of $A$, $B$, and all of the probe particles
would be given by
\begin{equation}
   |\mbox{final}\rangle = P_{M}U_{2}U_{1}|\Psi \rangle _{AB}\otimes
   |\Phi \rangle _{1}\otimes |\Phi \rangle _{2}.
\label{final}\end{equation}
Features $\alpha $ and $\beta $ of course follow from the expression for
$|\mbox{final}\rangle $ given in eq.\  \ref{final} (the presence of
$P_M$ being irrelevant); feature $\beta $ follows from the fact that
the projection operator $\mbox{\bf I}_{A}\otimes \mbox{\bf I}_{B}
\otimes |\Phi \rangle _{11}\langle \Phi |\otimes
(\mbox{\bf I}_{2}-|\Phi \rangle _{22}\langle \Phi |)$ annihilates
$|\mbox{final}\rangle$. (Note that the symbol {\bf I} denotes an
identity, not an isospin, operator.)

Features $\alpha $ and $\beta $ are features of the experimental results
which, according to standard quantum theory, would be obtained if an
experiment of the type discussed were actually performed.  To say that
an observation of $|\Phi \rangle _1$ prepares the $AB$
system in the isoscalar state is to make an interpretation of 
those features.
The HK prescription may lead to a different interpretation but, 
as will be shown below, to the same experimental features. So
let me consider this same example ---
interactions at $L1$, $L2$, $R1$, and $R2$, and the measurement
at $M$ --- as described by the HK prescription.  I first consider the case, 
illustrated in fig.\ 1, in which the points $R1$ and $R2$  both have
space-like separation from the measurement point $M$; this means,
in the HK prescription, that collapse due to that measurement has 
already occurred at the points $R1$ and $R2$. But that collapse has not 
occurred at $L1$ or $L2$, and so there is no single state-vector which
describes the system at $L1$ and at $R1$ (nor one for the system at $L2$ 
and $R2$).   Nevertheless, HK could proceed in the following way:
The unitary operator $U_1$ can (necessarily, since the
interactions are local) be written
\begin{equation}
U_{1} = (\mbox{\bf I}_{L}\otimes U_{R1})(U_{L1}\otimes \mbox{\bf I}_{R}),
\end{equation}
where $U_{R1}$ acts only on the particles present at $R1$, and $U_{L1}$
only on the particles present at $L1$.  Similarly,
$U_{2}=(\mbox{\bf I}_{L}\otimes U_{R2})(U_{L2}\otimes \mbox{\bf I}_{R})$.
The effect of the measurement at $M$ is to apply the projection
operator $P_M$ to the state vector, and to use the collapsed state at
$R1$ and $R2$ requires that  $P_M$ be
applied {\em before} the operators $U_{R1}$ or $U_{R2}$. So HK would
calculate
\begin{equation}
   |\mbox{final}\rangle =(\mbox{\bf I}_{L}\otimes U_{R2}) 
                         (\mbox{\bf I}_{L}\otimes U_{R1})P_{M}
                         (U_{L2}\otimes \mbox{\bf I}_{R})
                         (U_{L1}\otimes \mbox{\bf I}_{R})
                         |\Psi \rangle _{AB}\otimes
                         |\Phi \rangle _{1}\otimes |\Phi \rangle _{2}.
\label{HKfinal}\end{equation}
However, since $P_M$ commutes with $U_{R1}$ and with $U_{R2}$, and
$U_{R1}$ commutes with $U_{L2}$, the left-hand sides of eqs.\
\ref{final} and \ref{HKfinal} are equal.  Since, as noted above,
features $\alpha $ and $\beta $ are consequences of the form of
$|\mbox{final}\rangle$, we see that these features are valid for HK also.

Thus although the HK prescription does not assign a unique state-vector to
the $AB$ system at points $L2$ and $R2$, it does give correct calculated
values.  Perhaps this point could be clarified by a slight change in the
example.  Suppose that, as indicated in fig.\ 2, the points
$R1$ and $R2$ were unambiguously in the {\em future} of the measurement
point $M$, all other aspects of this example being unchanged. Then
one could still calculate $|\mbox{final}\rangle$
as in eq.\ \ref{HKfinal}, and  would see that features
$\alpha $ and $\beta $ were valid for this changed example also.
However, in this case not everybody would feel compelled to interpret
these features as implying that the $AB$ system at points $L2$ and $R2$
must be described by a unique state-vector.
\begin{figure}
\centering
\epsfig{figure=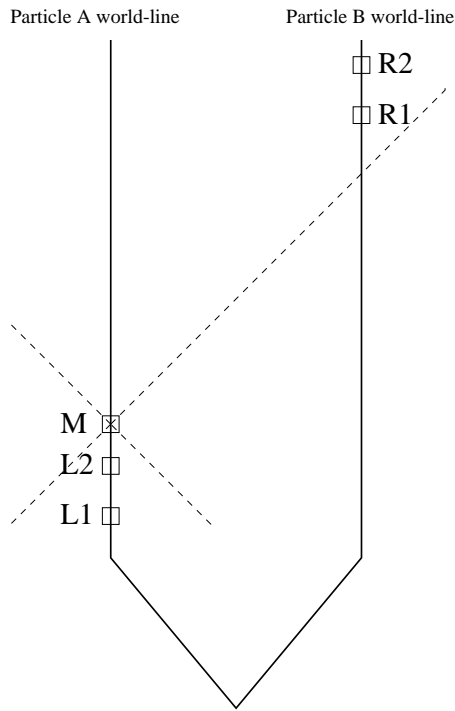}
\caption{The same as fig.\ 1, except that here $R1$ and $R2$ are
   unambiguously in the future of the measurement at $M$.}
\end{figure}

Of course one could construct more elaborate examples than the one
considered here, perhaps with
more than a single measurement-induced collapse. Any such example could
be analyzed, using the HK states, by the following procedure:
First, pick any single world-line (it might be convenient, but it is
not necessary, to pick the world-line of part of the quantum system
being described), and second, to use the HK states defined along that
world-line, which means to
apply the projection operators 
which collapse the state-vector in the order in which the backward
light cones of the measurements which produced those collapses
intersect the chosen world-line.  The HK prescription is constructed to 
give results
in complete agreement with standard quantum theory for any collection of
local measurements, and that remains true whether or not those 
measurements have the {\em effect}
(as in the example considered above) of measuring a non-local observable.

The HK prescription assigns to each space-time point a single 
state-vector,
and as I have argued above these state-vectors suffice to reproduce
the standard quantum result.  However, other prescriptions are also 
possible.
My personal preference (expressed in ref.\ \cite{me}; see also 
Mosley \cite{Mos}) is for the assumption that collapse occurs
along the {\em forward}  light cone of the measurement point. In this 
prescription, projection operators are applied in the order in which the
forward light cones of the measurements intersect the chosen world-line.

Aharonov and Albert \cite {AA3} have shown how a state-vector can be
assigned to each (unbounded) space-like surface.  Any such surface
separates the remainder of 
space-time into two disjoint regions, which may be called 
the regions ``before'' and ``after'' that surface (a point $M$ is
``before'' (``after'') the surface if the forward (backward) light cone
with vertex at $M$ intersects that surface);  Aharonov and Albert
assign to each space-like surface the state-vector which has experienced
collapses due to all those measurement interactions which are
``before'' the surface.  Suppose we accept this assignment, and suppose
that, for any space-time point $P$ we let $\eta (P)$ denote the 
forward light cone with vertex at $P$. We can then simply {\em define},
for each point $P$, the state $|\Psi (\eta (P))\rangle $ to be the
state assigned in ref.\ \cite{AA3} to the surface $\eta (P)$
(here I ignore, as do refs.\ \cite{G1,G3} in a similar context discussed
below, the fact that a light cone is not quite a space-like surface).
This definition, made entirely within the scheme of ref.\ \cite{AA3},
certainly does assign to each point $P$ a unique state-vector.
Furthermore, since for any space-time point $M$ the condition
``$M$ is before $\eta (P)$'' is equivalent to the condition
``$P$ is after the backward light cone from $M$'', it is
not hard to see that  $|\Psi (\eta (P))\rangle $ is exactly the state
which HK assign to the point $P$.  This indicates that the HK prescription,
rather than being in conflict with the prescription of Aharonov and Albert,
is really a special case of the latter.

Similarly, if we let $\sigma (P)$ denote the backward light cone with
vertex at $P$, we can define $|\Psi (\sigma (P))\rangle $ to be the
state which Aharonov and Albert
assign to the surface $\sigma (P)$.  It can then be seen
that $|\Psi (\sigma (P))\rangle $ is the same as the state assigned to
$P$ by the prescription (refs.\ \cite{me,Mos}) that collapse
occurs along the {\em forward} light cone of the measurement. So this 
prescription also can be considered as a special case of the prescription 
of ref.\ \cite{AA3}.

Ghirardi (\cite{G1,G3}) has suggested that this state-vector
$|\Psi (\sigma (P))\rangle $ should be used for the attribution of
local properties to a quantum system at point $P$ (for simplicity,
I ignore the fact that Ghirardi really modifies the surface
$\sigma (P)$ by the surface on which initial conditions are specified).
That is, he proposes that a system at the space-time point $P$ 
should be said to possess a definite value of a local quantity
just in case that  $|\Psi (\sigma (P))\rangle $ is an eigenvector of
the operator associated with that quantity (for brevity, let's read
that ``local properties of a system at $P$ are given by 
$|\Psi (\sigma (P))\rangle $'').
For the attribution of
non-local properties (such as the total isospin of a two-particle system)
at points $P$ and $P^{\prime }$, he defines a surface 
$\sigma (P,P^{\prime })$ as illustrated in fig.\ 3 (this is the surface
which lies immediately ``after'' the union of $\sigma (P)$ with
$\sigma (P^{\prime })$).  Defining 
$|\Psi (\sigma (P,P^{\prime })\rangle $ to be the state associated
with the surface $\sigma (P,P^{\prime })$, he proposes that non-local
properties be given by $|\Psi (\sigma (P,P^{\prime })\rangle $.

\begin{figure}
\centering
\epsfig{figure=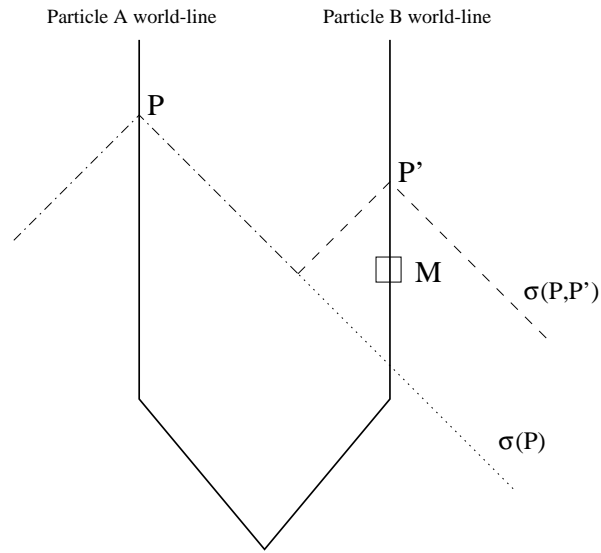}
\caption{Surfaces used for property attributions in ref.\ \cite{G1,G3}.
         Dashed lines represent $\sigma (P, P^{\prime})$, dotted line
         $\sigma(P)$, and dashed-dotted lines both surfaces coinciding.}
\end{figure}

This use of different surfaces for local and for non-local properties
can lead to some curious attributions, as is shown by the
following example. Suppose that  the initial state of a pair of
particles called $A$ and $B$ is a
superposition of a $\pi \pi $ and a $K\bar{K}$ pair. Since a $\pi$
has $I=1$ and a $K$ or $\bar{K}$ has $I=\frac{1}{2}$, I can write
\begin{eqnarray} |K\bar{K},I=0\rangle & = &
   \mbox{\small $\frac{1}{\sqrt{2}}$}(|K^{+}\rangle_ {A}|K^{-}\rangle _{B}
   -|K^{0}\rangle_ {A}|\bar{K}^{0}\rangle _{B}) \nonumber \\
    |\pi \pi , I=2,I_{z}=0\rangle & = &
 \mbox{\small $\frac{1}{\sqrt{6}}$}|\pi ^{+}\rangle _{A}|\pi ^{-}\rangle _{B}
+\mbox{\small $\sqrt{\frac{2}{3}}$}|\pi ^{0}\rangle _{A}|\pi ^{0}\rangle _{B}
+\mbox{\small $\frac{1}{\sqrt{6}}$}|\pi ^{-}\rangle _{A}|\pi ^{+}\rangle _{B}.
\nonumber
\end{eqnarray}
Now let me take the initial state of the $AB$ system to be
\begin{equation}
  |\Psi \rangle _{AB} = 
       \mbox{\small $\frac{1}{\sqrt{2}}$}(|K\bar{K},I=0\rangle 
       + |\pi \pi , I=2,I_{z}=0\rangle) ,
\end{equation}
and suppose that at the position marked $M$ in fig.\ 3 the particle
type but not the charge is measured (i.e., the hypercharge of particle $B$
is measured), and particle $B$ is found to be a $\pi$. Then since $M$ is
``before'' $\sigma (P,P^{\prime})$ but ``after'' $\sigma (P)$
we would have
\begin{equation} |\Psi(\sigma (P,P^{\prime}))\rangle =
                 |\pi \pi , I=2,I_{z}=0\rangle .
\end{equation}
but $|\Psi(\sigma (P ))\rangle $ is the same as 
the initial state $|\Psi \rangle _{AB}$.
Since $|\Psi(\sigma (P,P^{\prime }))\rangle $ is an eigenstate of $I^{2}$,
the proposed \cite{G1,G3} property attribution would be that  the combined 
state of $A$ at $P$ and $B$ at $P^{\prime} $
definitely has $I=2$, although the particle type of particle $A$ 
(which being a local property is determined by
$|\Psi(\sigma (P ))\rangle $ ) is not
definitely either $\pi$ or $K$. What is curious about this is that it
can be seen from the initial state that if $I=2$ then particle $A$
{\em is} definitely $\pi$.  In fact, independently of the chosen initial
state, a two-meson system with $I=2$ can never contain a $K$.

This curious situation would not occur if both local and non-local
properties  at $P$ were all given by
$|\Psi(\sigma (P ))\rangle $; in this 
example, that would mean that at point $P$ the value of $I$ would {\em not}
be definite.  The property attribution which is suggested by the 
prescription given in \cite{me,Mos} is that, {\em as seen from point $P$}, 
all properties are given by $|\Psi(\sigma (P ))\rangle $, whether or not
(any part of) the system being described happens to be located at $P$.
For local properties of a system located at $P$, this attribution
agrees with the one suggested in \cite{G1,G3}.

Finally, let me comment on two recent experimental results for which
the above considerations are relevant.
Scarani, Tittel, Zbinden and Gisin \cite{Gis1} report on an
experimental determination of a lower bound of $(1.5\times 10^{4})c$ for
what they call the
``speed of quantum information'' in the frame of the cosmic microwave
background radiation.  What is actually determined, as can be seen from
eq.\ 1 of \cite{Gis1}, is the distance between measurements performed on
a pair of entangled photons divided by the time difference of those
measurements.  It is not clear that this represents the speed of
anything, since it is not clear that there is anything which propagates
from one measurement to the other (e.g., according to the prescription
of refs.\ \cite{me}
and \cite{Mos}, not even state-vector collapse exceeds $c$).
In fact, footnote 14 of \cite{Gis1} suggests that, in spite of the title
of that paper, the ``speed of quantum information'' might not be the
most appropriate term for the quantity which that experiment determined;
I would agree with that suggestion.

Zbinden, Brendel, Gisin and Tittel
\cite{Gis2} report on an attempt to find deviations from 
standard quantum predictions in a case in which the detectors 
for two entangled photons are in relative motion in such a way that each
detector acts first in its own reference frame.  No such deviations were
found.  As discussed above, the HK prescription for state-vector collapse is
completely consistent with standard quantum predictions (as are the 
prescriptions of ref.\ \cite{AA3} and of refs.\ \cite{me,Mos}).
Furthermore, the CSL model (reviewed in refs.\ \cite{G1,G2})
would presumably give results which would be indistinguishable from
the standard predictions to this level of precision.
This casts doubt on the assertion made in the last
sentence of \cite{Gis2} that experimental results which agree with standard
quantum theory ``make it more difficult to view the `projection postulate'
as a compact description of a real physical phenomenon''.

\section*{Acknowledgements}
I acknowledge the hospitality of the
Lawrence Berkeley National Laboratory.

\end{document}